\documentstyle[12pt]{article}
\begin{document}

\title{\Large\bf Functional versus canonical quantization of a
nonlocal massive vector-gauge theory}

\author{R. Amorim\thanks{\noindent e-mail: amorim@if.ufrj.br}~~ and
J. Barcelos-Neto\thanks{\noindent e-mail: barcelos@if.ufrj.br}\\ 
Instituto de F\'{\i}sica\\ 
Universidade Federal do Rio de Janeiro\\ 
RJ 21945-970 - Caixa Postal 68528 - Brasil}
\date{}

\maketitle
\abstract
It has been shown in literature that a possible mechanism of mass
generation for gauge fields is through a topological coupling of
vector and tensor fields.  After integrating over the tensor degrees
of freedom, one arrives at an effective massive theory that, although
gauge invariant, is nonlocal. Here we quantize this nonlocal
resulting theory both by path integral and canonical procedures. This
system can be considered as equivalent to one with an infinite number
of time derivatives and consequently an infinite number of momenta.
This means that the use of the canonical formalism deserves some
care. We show the consistency of the formalism we use in the
canonical procedure by showing that the obtained propagators are the
same as those of the (Lagrangian) path integral approach. The problem
of nonlocality appears in the obtainment of the spectrum of the
theory.  This fact becomes very transparent when we list the infinite
number of commutators involving the fields and their velocities.

\vfill
\vspace{1cm}
\newpage

\section{Introduction}

\bigskip
It is widely accepted that the forces of nature are described by
gauge theories. These theories are characterized by the gauge
symmetries which are related to massless fields. However, sometimes
it is necessary that these fields become massful, as it occurs, for
instance, in the case of the Salam-Weinberg theory.  Nowadays, it has
also been widely accepted that spontaneous symmetry breaking together
with the Higgs mechanism are the most probable explanation for the
origin of the acquisition of mass by gauge fields. However, if this
is actually true, the Higgs bosons must exist in nature. The point is
that there is no precise theoretical prediction on the mass scale
where these fields could be found and experiments till now have shown
no evidence about them.

\medskip
In this way, alternative mechanisms of mass generation for gauge
fields that do not spoil what is well established and that do not
contain Higgs bosons are welcome. This might be the case of
vector-tensor gauge theories \cite{Allen}, where vector and tensor
fields are coupled in a topological way by means of a kind of
Chern-Simons term. The general idea of this mechanism resides in the
following: tensor gauge fields \cite{Kalb} are antisymmetric
quantities and consequently in $D=4$ they exhibit six degrees of
freedom. By virtue of the massless condition, the number of degrees of
freedom goes down to four. Since the gauge parameter is a vector
quantity, this number would be zero if all of its components were
independent. This is nonetheless the case because the system is
reducible (which means that the gauge transformations are not all
independent) and we mention that the final number of physical degrees
of freedom is one. It is precisely this degree of freedom that can be
absorbed by the vector gauge field in the vector-tensor gauge theory
in order to acquire mass \cite{Allen,Barc1}. This peculiar structure
of constraints involving tensor gauge theories implies that
quantization as well as its non-Abelian formulation deserve some care
and a reasonable amount of work has been done on these subjects
\cite{Kaul,Henneaux,Freedam, Barc2}.

\medskip
Usually, the treatment of the vector-tensor gauge theory is carried
out with both vector and tensor fields placed together and just at
the end the integration over the tensor field is done in order to
obtain the effective result for the vector theory. This procedure
usually hides an important aspect of this effective theory, that is
its nonlocality. We mention that it is equivalent to a theory with an
infinite number of higher derivative terms and, consequently, an
infinite number of momenta also.

\medskip
It is not our purpose here to advocate if a vector-tensor gauge
theory with topological coupling is more suitable to explain the mass
generation than the usual spontaneous symmetry breaking together with
the Higgs mechanism. Our intention in the present paper is to study
the quantization of massive vector gauge field directly by means of
the nonlocal effective Lagrangian \cite{Marino}. We do it both by
path integral, where we use a Lagrangian formulation, as well as by
the canonical approach. This is the subject of Sections 2 and 3
respectively. We would like also to add that the non-Abelian
formulation for the vector tensor gauge theory is not a simple task.
This is so because the non-Abelian version loses the reducibility
condition unless we consider that the Maxwell stress tensor as zero
\cite{Freedam}.  Another possibility is to introduce a kind of
Stuckelberg field, that disappears in the Abelian limit, in order to
keep the same number of degrees of freedom in both sectors (Abelian
and non-Abelian) of the theory \cite{Barc2}. We shall consider only
the Abelian case in this paper.  We left Sec. 4 for some concluding
remarks and include four appendices to present details of some
calculations.

\vspace{1cm}
\section{Brief review of the vector-tensor gauge theory and the path
integral quantization of the effective vector theory} 
\renewcommand{\theequation}{2.\arabic{equation}}
\setcounter{equation}{0}

\bigskip
The Abelian theory for vector and tensor fields coupled in a
topological way is described by the Lagrangian density \cite{Allen}:

\begin{equation}
{\cal L}=-\,\frac{1}{4}\,F_{\mu\nu}\,F^{\mu\nu}
+\,\frac{1}{12}\,H_{\mu\nu\rho}\,H^{\mu\nu\rho}\,
+{m\over2}\,\epsilon^{\mu\nu\rho\lambda}\,
B_{\mu\nu}F_{\rho\lambda}\,,
\label{2.1}
\end{equation}

\bigskip\noindent 
where $F_{\mu\nu}$ and $H_{\mu\nu\rho}$ are totally antisymmetric
tensors written in terms of the potentials $A_{\mu}$ and $B_{\mu\nu}$
(also antisymmetric) through the stress tensors

\begin{eqnarray}
F_{\mu\nu}&=&\partial_\mu A_{\nu}-\partial_\nu A_{\mu}\,,
\label{2.2}\\
H_{\mu\nu\rho}&=&\partial_\mu B_{\nu\rho}
+\partial_\rho B_{\mu\nu}+\partial_\nu B_{\rho\mu}\,.
\label{2.3}
\end{eqnarray}

\bigskip\noindent
In expression (\ref{2.1}), $\epsilon^{\mu\nu\rho\lambda}$ is the
totally antisymmetric symbol and $m$ is a mass parameter. It is easy
to see, by using the (coupled) Euler-Lagrange equations for $A^\mu$
and $B^{\mu\nu}$, as well as the Jacobi identity, that $F_{\mu\nu}$
satisfy a massive Klein-Gordon equation, with a mass parameter
$m$~\cite{Allen}.

\medskip
We observe that the Lagrangian (\ref{2.1}) is invariant under the
gauge transformations

\begin{eqnarray}
\delta A^\mu&=&\partial^\mu\Lambda\,,
\label{2.4}\\
\delta B^{\mu\nu}&=&\partial^\mu\Lambda^\nu
-\partial^\nu\Lambda^\mu\,,
\label{2.5}
\end{eqnarray}

\bigskip\noindent
where $\Lambda$ and $\Lambda^\mu$ are (before fixing the gauge)
generic functions of spacetime. This is a reducible theory, what
means that not all the gauge transformations above are independent.
In fact, if we choose the gauge parameter $\Lambda^\mu$ as the
gradient of some scalar $\Omega$ we have that $B^{\mu\nu}$ does not
change under the gauge transformation (\ref{2.5}).

\bigskip
Functionally integrating over the antisymmetric tensor field
$B_{\mu\nu}$ we get, after a convenient gauge fixing procedure, the
effective action \cite{Barc1}

\begin{equation}
S_0\,[A_\mu]=-\,\frac{1}{4}\int d^4x\,F_{\mu\nu}\,
\Bigl(1+\frac{m^2}{\Box}\Bigr)\,F^{\mu\nu}\,.
\label{2.6}
\end{equation}

\bigskip\noindent 
The action (\ref{2.6}), although nonlocal, is gauge invariant. It is
important to emphasize that this theory is renormalizable, a
characteristic that is lost when a mass term is directly put by hand
as in the Proca theory. 

\medskip
Let us calculate the (covariant) propagator for the field $A_\mu$. We
opt to use a Lagrangian formulation in order to avoid the problem of
the infinite number of momenta. Let us use the Batalin-Vilkovisky
(BV) formalism \cite{Henneaux,BV}. The nonminimum BV action can be
written as

\begin{equation}
S=S_0 + \int dx\,\bigl( A^\ast_\mu\,\partial^\mu c 
+b\,\bar c^\ast\bigr)\,,
\label{2.7}
\end{equation}

\bigskip\noindent 
where the $A^\ast_\mu$ and the pair ($c^\ast,\bar c^\ast$)  are
respectively the antifields of the gauge field $A^\mu$ and of the
ghosts ($c,\bar c$). The auxiliary field $b$ was introduced in order
to fix the gauge in a covariant way. This can be done, for instance,
with the aid of the gauge-fixing fermion functional

\begin{equation}
\Psi=\int dx\,\bar c\,\bigl(-\,\alpha b
+\partial^\mu A_\mu\bigr)\,,
\label{2.8}
\end{equation}

\bigskip\noindent 
with $\alpha$ being a parameter. The vacuum functional is defined by
\cite{Henneaux,BV} 

\begin{equation}
Z_\Psi=\int\,[dA_\mu][d\bar c][dc][db]
[dA^\ast_\mu][d\bar c^\ast][dc^\ast]\,
\delta\Bigl[\phi^\ast-\frac{\delta\Psi}{\delta\phi}\Bigr]\,
\exp\,\{iS\}\,.
\label{2.9}
\end{equation}

\bigskip\noindent 
The action $S$ is given by (\ref{2.7}) and $\phi$ is generically
referring to gauge and ghost fields. After functionally integrating
over the antifields as well as over the auxiliary field $b$, we
arrive at

\begin{equation}
\bar S\,[J]=\int\,d^4x\,\Bigl[-\frac{1}{4}\,F_{\mu\nu}\,
\Bigl(1+\frac{m^2}{\Box}\Bigr)\,F^{\mu\nu}
-\partial_\mu\bar c\,\partial^\mu c
+\frac{1}{2\alpha}\,\Bigl(\partial^\mu A_\mu\Bigr)^2
+J_\mu A^\mu\Bigr]\,,
\label{2.10}
\end{equation}

\bigskip\noindent 
where we have introduced an external source $J^\mu$ in order to
calculate the propagator. This can be directly obtained by a
straightforward calculation. The result, written in momentum space,
is 

\begin{equation}
K^{\mu\nu}=-\,\frac{1}{k^2+m^2}\,
\Bigl[\eta^{\mu\nu}+\Bigl(\frac{\alpha-1}{k^2}
+\frac{m^2}{k^4}\Bigr)\,k^\mu k^\nu\Bigr]\,.
\label{2.11}
\end{equation}

\bigskip\noindent 
We notice that there is actually a mass pole at $k_0^2=\vec k^2+m^2$.

\medskip
In the next Section we are going to see how this and other features
appear in terms of a canonical quantization procedure.

\vspace{1cm}
\section{Canonical quantization} 
\renewcommand{\theequation}{3.\arabic{equation}}
\setcounter{equation}{0}

\bigskip
Let us consider the Lagrangian for vector fields of Eq. (\ref{2.10})

\begin{equation}
{\cal L}=-\frac{1}{4}\,F_{\mu\nu}\,
\Bigl(1+\frac{m^2}{\Box}\Bigr)\,F^{\mu\nu}
-\frac{1}{2\alpha}\,\Bigl(\partial^\mu A_\mu\Bigr)^2\,.
\label{3.1}
\end{equation}

\bigskip\noindent
To implement the process of canonical quantization for such system,
it is necessary to obtain the canonical momenta. So, we have to
isolate the time derivatives from the nonlocal operator $\Box^{-1}$.
We then conveniently write

\begin{eqnarray}
\frac{1}{\Box}&=&-\,\bigl(\nabla^2-\partial_t^2\bigr)^{-1}\,,
\nonumber\\
&=&-\frac{1}{\nabla^2}
-\frac{\partial_t^2}{\nabla^4}
-\frac{\partial_t^4}{\nabla^6}-\cdots
\label{3.2}
\end{eqnarray}

\bigskip
As one observes, the system described by (\ref{3.1}) is effectively a
system with an infinite number of time derivatives and consequently
it contains an infinite number of momenta \cite{High,Barc3}.  A
practical way of obtaining the momentum expressions is to consider the
variation of the action by fixing the fields and their velocities at
just one of the extreme times, say, $\delta A_\mu(\vec
x,t_0)=0=\delta\dot A_\mu(\vec x,t_0)=\delta\ddot A_\mu(\vec
x,t_0)=\cdots$. After some algebraic calculation, we get (please, see
Appendix A)

\begin{eqnarray}
\delta\int_{t_0}^t d\tau\int d^3\vec x\,\,{\cal L}
&=&\int_{t_0}^t d\tau\int d^3\vec x\,\biggl[
\Big(1+\frac{m^2}{\Box}\Bigr)\,\partial^\mu F_{\mu\nu}
+\frac{1}{\alpha}\,\partial_\mu\partial_\nu A^\mu\biggr]\,
\delta A^\nu
\nonumber\\
&&-\int d^3\vec x\,\biggl[\Bigl(1+\frac{m^2}{\Box}\Bigr)\,F_{0\nu}
+\frac{m^2}{2\Box}\,\frac{\partial^i}{\nabla^2}\,\dot F_{i\nu}
+\frac{1}{\alpha}\,\partial_\mu A^\mu\,\eta_{0\nu}\biggl]\,
\delta A^\nu
\nonumber\\
&&+\,\frac{m^2}{2}\int d^3\vec x\,\,\frac{1}{\Box}\,
\frac{\partial^\mu}{\nabla^2}\,F_{\mu\nu}\,
\delta\dot A^\nu
\nonumber\\
&&-\,\frac{m^2}{2}\int d^3\vec x\,\,\frac{1}{\Box}\,
\Bigl(\frac{1}{\nabla^2}\,F_{0\nu}
+\frac{\partial^i}{\nabla^4}\,\dot F_{i\nu}\Bigr)\,
\delta\ddot A^\nu
\nonumber\\
&&+\,\frac{m^2}{2}\int d^3\vec x\,\,\frac{1}{\Box}\,
\frac{\partial^\mu}{\nabla^4}\,F_{\mu\nu}\,
\delta\buildrel{...}\over A^\nu
\nonumber\\
&&+\,\cdots
\label{3.3}
\end{eqnarray}

\bigskip\noindent
The coefficient of $\delta A^\nu$ in the first term is the equation
of motion, namely,

\begin{equation}
\Big(1+\frac{m^2}{\Box}\Bigr)\,\partial^\mu F_{\mu\nu}
+\frac{1}{\alpha}\,\partial_\mu\partial_\nu A^\mu=0\,.
\label{3.4}
\end{equation}

\bigskip\noindent
In the remaining terms, the coefficients of $\delta A^\nu$,
$\delta\dot A^\nu$, $\delta\ddot A^\nu$ etc. are the canonical
momentum conjugate to $A^\nu$, $\dot A^\nu$, $\ddot A^\nu$ etc.
Denoting these momenta by $\pi_\nu$, $\pi_\nu^{(1)}$, $\pi_\nu^{(2)}$
etc., we have

\begin{eqnarray}
\pi_\nu&=&-\,\Bigl(1+\frac{m^2}{\Box}\Bigr)\,F_{0\nu}
-\frac{m^2}{2\Box}\frac{\partial^i}{\nabla^2}\,\dot F_{i\nu}
-\frac{1}{\alpha}\,\partial_\mu A^\mu\,\eta_{0\nu}\,,
\nonumber\\
\pi_\nu^{(1)}&=&\frac{m^2}{2\Box}\,
\frac{\partial^\mu}{\nabla^2}\,F_{\mu\nu}\,,
\nonumber\\
\pi_\nu^{(2)}&=&-\frac{m^2}{2\Box}\,\frac{1}{\nabla^2}\,
\Bigl(F_{0\nu}+\frac{\partial^i}{\nabla^2}\,\dot F_{i\nu}\Bigr)\,,
\nonumber\\
\pi_\nu^{(3)}&=&\frac{m^2}{2\Box}\,
\frac{\partial^\mu}{\nabla^4}\,F_{\mu\nu}\,,
\nonumber\\
\pi_\nu^{(4)}&=&-\,\frac{m^2}{2\Box}\,\frac{1}{\nabla^4}
\Bigl(F_{0\nu}+\frac{\partial^i}{\nabla^2}\,\dot F_{i\nu}\Bigr)\,,
\nonumber\\
&\vdots&
\label{3.5}
\end{eqnarray}

\bigskip
Systems with higher derivatives have fields and their velocities as
independent coordinates. For example, a system with two derivatives
has its fields (denoting them generically by $\phi$) and their
velocities $\dot\phi$ as independent coordinates. If there are no
constraints in the theory, the Poisson brackets (P.B.) are the bridge
to the quantum commutator. Thus, we must have $[\phi,\dot\phi]=0$.
The commutators that might not be zero are those (in this example
with two derivatives) involving $\phi$ and $\dot\phi$ with the higher
derivatives $\ddot\phi$ and $\buildrel{\rm\bf...}\over\phi$
\cite{Barc3}. 

\medskip
The problem that comes out in the system we are studying is that
there is an infinite number of time derivatives and it is not clear a
priori which commutators are not zero. In order to try to figure them
out we take the Lagrangian (\ref{3.1}), expanded in $t$ derivatives,
till a certain limit order $n$ and at the end we let $n$ goes to
infinity. Let us then consider the expansion (\ref{3.2}) till
$\partial_t^2$, which is the first nontrivial order,

\begin{equation}
{\cal L}_3=-\,\frac{1}{4}\,F_{\mu\nu}F^{\mu\nu}
+\frac{m^2}{4}\,F_{\mu\nu}\,
\Bigl(1+\frac{\partial_t^2}{\nabla^2}\Bigr)\,
\frac{1}{\nabla^2}\,F^{\mu\nu}
-\,\frac{1}{2\alpha}\,\bigl(\partial_\mu A^\mu\bigr)^2\,.
\label{3.7}
\end{equation}

\bigskip\noindent
The equation of motion and the momenta are given by

\begin{eqnarray}
&&\partial^\mu F_{\mu\nu}
-m^2\,\Bigl(1+\frac{\partial_t^2}{\nabla^2}\Bigr)\,
\frac{\partial^\mu}{\nabla^2}\,F_{\mu\nu}
+\frac{1}{\alpha}\,\partial_\mu\partial_\nu A^\mu=0\,,
\label{3.8}\\
&&\pi_\nu=-\,F_{0\nu}
+m^2\,\Bigl(1+\frac{\partial_t^2}{\nabla^2}\Bigr)\,
\frac{1}{\nabla^2}\,F_{0\nu}
+\frac{m^2}{2}\,\frac{\partial^i\partial_t}{\nabla^4}\,F_{i\nu}
-\frac{1}{\alpha}\,\eta_{0\nu}\,\partial_\mu A^\mu\,,
\label{3.9}\\
&&\pi_\nu^{(1)}=-\,\frac{m^2}{2}\,
\frac{\partial^\mu}{\nabla^4}\,F_{\mu\nu}\,,
\label{3.10}\\
&&\pi_\nu^{(2)}=\frac{m^2}{2}\,\frac{1}{\nabla^4}\,F_{0\nu}\,.
\label{3.11}
\end{eqnarray}

\bigskip\noindent
In this order, the phase-space coordinate is given by
$(A_\mu,\pi^\nu)$ $\oplus$ $(\dot A_\mu,\pi^{(1)\nu})$ $\oplus$
$(\ddot A_\mu,\pi^{(2)\nu})$. We thus observe that relations
(\ref{3.10}) and (\ref{3.11}) are constraints, as well as the zero
component of $\pi_\nu$. The fundamental nonvanishing P.B. are

\begin{eqnarray}
\bigl\{A_\mu(\vec x,t),\pi^\nu(\vec y,t)\bigr\}
&=&\delta_\mu^\nu\,\delta(\vec x-\vec y)\,,
\nonumber\\
\bigl\{\dot A_\mu(\vec x,t),\pi^{(1)\nu}(\vec y,t)\bigr\}
&=&\delta_\mu^\nu\,\delta(\vec x-\vec y)\,,
\nonumber\\
\bigl\{\ddot A_\mu(\vec x,t),\pi^{(2)\nu}(\vec y,t)\bigr\}
&=&\delta_\mu^\nu\,\delta(\vec x-\vec y)\,.
\label{3.12}
\end{eqnarray}

\bigskip\noindent
In order to calculate the P.B. matrix of the constraints, it is
convenient to develop them separating all the velocities. The
result is

\begin{eqnarray}
T_0&=&\pi_0+\Bigl(\frac{1}{\alpha}
+\frac{m^2}{2\nabla^2}\Bigr)\,\dot A_0
+\frac{m^2}{2\nabla^4}\,\partial^i\ddot A_i
+\frac{1}{\alpha}\,\partial_i A^i\,,
\label{3.13}\\
T_\nu^{(1)}&=&\pi_\nu^{(1)}
-\frac{m^2}{2\nabla^4}\Bigl[\nabla^2\,A_\nu
+\delta_\nu^0\partial_i\dot A^i
-\delta_\nu^i(\ddot A_i-\partial_i\dot A_0
-\partial_i\partial_j A^j)\Bigr]\,,
\label{3.14}\\
T_\nu^{(2)}&=&\pi_\nu^{(2)}
+\delta_\nu^i\,\frac{m^2}{2\nabla^4}\,
\Bigl(\partial_i A_0-\dot A_i\Bigr)\,.
\label{3.15}
\end{eqnarray}

\bigskip\noindent
We observe that the last constraint for $\nu=0$ becomes
$T_0^{(2)}=\pi_0^{(2)}$. We also observe that the other constraints
do not contain $\ddot A_0$ and consequently the P.B. matrix for the
constraints above will be singular (in fact, $\ddot A_0$ does not
play any role in the theory). Thus, instead of the constraint
(\ref{3.15}), we take

\begin{equation}
T_i^{(2)}=\pi_i^{(2)}
+\frac{m^2}{2\nabla^4}\,
\Bigl(\partial_i A_0-\dot A_i\Bigr)\,.
\label{3.16}
\end{equation}

\bigskip\noindent
The P.B. matrix of the constraints reads

\begin{equation}
S=\left(\begin{array}{ccc}
0&m^2\delta_0^\nu\Bigl(\frac{1}{\alpha}+\frac{1}{\nabla^2}\Bigr)
&m^2\frac{\partial^j}{\nabla^4}\\
&&\\
-\,m^2\delta^0_\mu\Bigl(\frac{1}{\alpha}+\frac{1}{\nabla^2}\Bigr)
&-\,m^2\Bigl(\delta_\mu^0\eta^{k\nu}
+\delta_\mu^k\eta^{0\nu}\Bigr)\frac{\partial_k}{\nabla^4}
&m^2\delta_\mu^j\frac{1}{\nabla^4}\\
&&\\
m^2\frac{\partial_i}{\nabla^4}
&-\,m^2\delta_i^\nu\frac{1}{\nabla^4}&0\\
\end{array}\right)\,\delta(\vec x-\vec y)\,.
\label{3.17}
\end{equation}

\bigskip
Since this matrix involves space and time indices separately, the
calculation of its inverse requires some care. The details of the
calculation is presented in Appendix B. The result reads

\begin{equation}
S^{-1}=\left(\begin{array}{ccc}
0&-\,\alpha\,\delta_0^\nu&\alpha\,\partial^j\\
&&\\
\alpha\,\delta_\mu^0
&-\,\alpha(\delta_\mu^0\delta_k^\nu\partial^k
+\delta_\mu^k\delta_0^\nu\partial_k)
&-\,\delta_\mu^k\Bigl(\delta_k^j\frac{\nabla^4}{m^2}
-\alpha\partial_k\partial^j\Bigr)\\
&&\\
\alpha\,\partial_i
&\delta^\nu_k\Bigl(\delta^k_i\frac{\nabla^4}{m^2}
-\alpha\partial^k\partial_i\Bigr)&0\\
\end{array}\right)\,\delta(\vec x-\vec y)\,.
\label{3.18}
\end{equation}

\bigskip\noindent
With this inverse, we directly obtain the following Dirac brackets
(D.B.) \cite{Dirac}

\begin{eqnarray}
\bigl\{\dot A_\mu(\vec x,t),A^\nu(\vec y,t)\bigr\}_D
&=&\alpha\,\delta_\mu^0\delta_0^\nu\,\delta(\vec x-\vec y)\,,
\nonumber\\
\bigl\{\ddot A_\mu(\vec x,t),A^\nu(\vec y,t)\bigr\}_D
&=&\alpha\,\delta_\mu^i\delta_0^\nu\partial_i\,
\delta(\vec x-\vec y)\,.
\label{3.19}
\end{eqnarray}

\bigskip\noindent
The bracket $\{\buildrel{...}\over A_\mu,A^\nu\}$ is obtained
from $\{\pi_\mu,A^\nu\}$. The result is

\begin{equation}
\{\buildrel{...}\over A_\mu(\vec x,t),A^\nu(\vec y,t)\}
=-\,\delta^\nu_\mu\,\frac{\nabla^4}{m^2}\,\delta(\vec x-\vec y)\,.
\label{3.20}
\end{equation}

\bigskip\noindent
The commutators follow directly from expressions (\ref{3.19}) and
(\ref{3.20}), i.e. 

\begin{eqnarray}
\bigl[\dot A_\mu(\vec x,t),A^\nu(\vec y,t)\bigr]
&=&i\,\alpha\,\delta_\mu^0\delta_0^\nu\,\delta(\vec x-\vec y)\,,
\nonumber\\
\bigl[\ddot A_\mu(\vec x,t),A^\nu(\vec y,t)\bigr]
&=&i\,\alpha\,\delta_\mu^i\delta_0^\nu\partial_i\,
\delta(\vec x-\vec y)\,,
\nonumber\\
\bigl[\buildrel{...}\over A_\mu(\vec x,t),A^\nu(\vec y,t)\bigr]
&=&-\,i\,\delta^\nu_\mu\,\frac{\nabla^4}{m^2}\,
\delta(\vec x-\vec y)\,.
\label{3.21}
\end{eqnarray}

\bigskip
It might be opportune and instructive to call our attention for the
following fact. We notice that from the first commutator above we get
$[\dot A_i(\vec x,t),A^j(\vec y,t)]=0$. Since there is no dependence
of this result with the mass parameter $m$, it may appear that there
is a conflict with the limit case when $m\rightarrow0$, where the
Maxwell theory is obtained. We know that this commutator is not zero
in the Maxwell theory. What happens it that in the limit of
$m\rightarrow0$, the structure of constraints is not the same as in
the massful case, and consequently, the results we obtain in one
sector cannot be kept in the other. We find important to explain this
point with details in order to reinforce the formalism we are using.
We use the Appendix C to do this.

\medskip
Let us now consider the propagator calculation. One can directly
show by using the path integral formalism that the propagator
corresponds to the inverse of the operator that appears in the
equation of motion. Considering (\ref{3.8}), we have

\begin{equation}
\biggl\{\eta_{\mu\nu}\biggl[1
-\Bigl(1+\frac{\partial_t^2}{\nabla^2}\Bigr)\,
\frac{m^2}{\nabla^2}\biggr]\Box
+\biggl[\frac{1}{\alpha}-1
+\Bigl(1+\frac{\partial_t^2}{\nabla^2}\Bigr)\,
\frac{m^2}{\nabla^2}\biggr]\,
\partial_\mu\partial_\nu\biggr\}\,A^\nu=0\,.
\label{3.22}
\end{equation}

\bigskip\noindent
This means that the propagator must satisfies the equation

\begin{eqnarray}
&&\biggl\{\eta_{\mu\nu}\biggl[1
-\Bigl(1+\frac{\partial_t^2}{\nabla^2}\Bigr)\,
\frac{m^2}{\nabla^2}\biggr]\Box
+\biggl[\frac{1}{\alpha}-1
+\Bigl(1+\frac{\partial_t^2}{\nabla^2}\Bigr)\,
\frac{m^2}{\nabla^2}\biggr]\,
\partial_\mu\partial_\nu\biggr\}
\nonumber\\
&&\phantom{\hspace{3cm}}
\times\,T(A^\nu(x)A^\rho(x^\prime))
=i\,\delta_\mu^\rho\,\delta(\vec x-\vec x^\prime)\,.
\label{3.23}
\end{eqnarray}

\bigskip\noindent
If what we have done till now is consistent, that is to say, if the
quantization embodied in the commutators (\ref{3.21}), the expression
(\ref{3.23}) ought to be verified. In fact, after a hard algebraic
calculation, we show that this actually occurs (see Appendix D for
some details).

\bigskip
Let us now consider the Lagrangian with the next term of the
expansion of $\Box^{-1}$, 

\begin{equation}
{\cal L}_4=-\,\frac{1}{4}\,F_{\mu\nu}F^{\mu\nu}
+\frac{m^2}{4}\,F_{\mu\nu}\Bigl(1
+\frac{\partial_t^2}{\nabla^2}
+\frac{\partial_t^4}{\nabla^4}\Bigr)\,
\frac{1}{\nabla^2}\,F^{\mu\nu}
-\frac{1}{2\alpha}\,\Bigl(\partial_\mu A^\mu\Bigr)^2\,.
\label{3.24}
\end{equation}

\bigskip\noindent
Proceeding as before we obtain the equation of motion 

\begin{equation}
\partial^\mu F_{\mu\nu}
-m^2\,\Bigl(1+\frac{\partial_t^2}{\nabla^2}
+\frac{\partial_t^4}{\nabla^4}\Bigr)\,
\frac{\partial^\mu}{\nabla^2}\,F_{\mu\nu}
+\frac{1}{\alpha}\,\partial_\mu\partial_\nu A^\mu=0
\label{3.25}
\end{equation}

\bigskip\noindent
and the momentum expressions

\begin{eqnarray}
\pi_\nu&=&-\,F_{0\nu}
+m^2\,\Bigl(1+\frac{\partial_t^2}{\nabla^2}
+\frac{\partial_t^4}{\nabla^4}\Bigr)\,
\frac{1}{\nabla^2}\,F_{0\nu}
\nonumber\\
&&+\,\frac{m^2}{2}\,\Bigl(1+\frac{\partial_t^2}{\nabla^2}\Bigr)\,
\frac{\partial^i\partial_t}{\nabla^4}\,F_{i\nu}
-\frac{1}{\alpha}\,\eta_{0\nu}\partial_\mu A^\mu\,,
\nonumber\\
\pi_\nu^{(1)}&=&-\,\frac{m^2}{2}\,
\Bigl(1+\frac{\partial^2_t}{\nabla^2}\Bigr)\,
\frac{\partial^\mu}{\nabla^4}\,F_{\mu\nu}\,,
\nonumber\\
\pi_\nu^{(2)}&=&\frac{m^2}{2}\,
\Bigl(1+\frac{\partial^2_t}{\nabla^2}\Bigr)\,
\frac{1}{\nabla^4}\,F_{0\nu}
+\frac{m^2}{2}\,\frac{\partial_t\partial^i}{\nabla^6}\,F_{i\nu}\,,
\nonumber\\
\pi_\nu^{(3)}&=&-\,\frac{m^2}{2\nabla^6}\,
\partial^\mu F_{\mu\nu}\,,
\nonumber\\
\pi_\nu^{(4)}&=&\frac{m^2}{2\nabla^6}\,F_{0\nu}\,.
\label{3.26}
\end{eqnarray}

\bigskip
The set of independent constraints is now given by

\begin{eqnarray}
T_0&=&\pi_0+\frac{1}{\alpha}\,\partial_iA^i
+\Bigl(\frac{m^2}{2\nabla^2}+\frac{1}{\alpha}\Bigr)\,\dot A_0
+\frac{m^2}{2}\,\frac{\partial^i}{\nabla^4}\,\ddot A_i
\nonumber\\
&&\phantom{\pi_0}
+\,\frac{m^2}{2\nabla^4}\,\buildrel{...}\over A_0
+\,\frac{m^2}{2}\frac{\partial^i}{\nabla^6}\,
\buildrel{....}\over A_i\,,
\nonumber\\
T_\mu^{(1)}&=&\pi_\mu^{(1)}
-\frac{m^2}{2\nabla^2}\,\Bigl(A_\mu+\delta_\mu^i\,
\frac{\partial_i\partial_j}{\nabla^2}\,A^j\Bigr)
-\frac{m^2}{2}\,\delta_\mu^0\,
\frac{\partial^i}{\nabla^4}\,\dot A_i
\nonumber\\
&&\phantom{\pi_\mu^{(1)}}
-\,\frac{m^2}{2}\,\delta_\mu^i\,
\frac{\partial_i}{\nabla^4}\,\dot A_0
-\frac{m^2}{2}\,\delta_\mu^0\,
\frac{1}{\nabla^4}\,\ddot A_0
-\frac{m^2}{2}\,\delta_\mu^i\,
\frac{\partial_i\partial_j}{\nabla^6}\,\ddot A^j
\nonumber\\
&&\phantom{\pi_\mu^{(1)}}
-\frac{m^2}{2}\,\delta_\mu^0\,
\frac{\partial^i}{\nabla^6}\,\buildrel{...}\over A_i
-\frac{m^2}{2}\,\delta_\mu^i\,
\frac{\partial_i}{\nabla^6}\,\buildrel{...}\over A_0
+\frac{m^2}{2}\,\delta_\mu^i\,
\frac{1}{\nabla^6}\,\buildrel{....}\over A_i\,,
\nonumber\\
T_i^{(2)}&=&\pi_i^{(2)}
+\frac{m^2}{2}\,\frac{\partial_i}{\nabla^4}\,A_0
+\frac{m^2}{2}\,\frac{\partial_i\partial_j}{\nabla^6}\,\dot A^j
\nonumber\\
&&\phantom{\pi_\mu^{(1)}}
+\,\frac{m^2}{2}\,\frac{\partial_i}{\nabla^4}\,\ddot A_0
-\frac{m^2}{2\nabla^6}\,\buildrel{...}\over A_i\,,
\nonumber\\
T_i^{(3)}&=&\pi_i^{(3)}
-\frac{m^2}{2\nabla^4}\,A_i
-\frac{m^2}{2}\,\frac{\partial_i\partial_j}{\nabla^6}\,A^j
\nonumber\\
&&\phantom{\pi_\mu^{(1)}}
-\frac{m^2}{2}\,\frac{\partial_i}{\nabla^6}\,\dot A_0
+\frac{m^2}{2\nabla^6}\,\ddot A_i\,,
\nonumber\\
T_i^{(4)}&=&\pi_i^{(4)}
+\frac{m^2}{2}\,\frac{\partial_i}{\nabla^6}\,A_0
-\frac{m^2}{2\nabla^6}\,\dot A^i\,,
\label{3.27}
\end{eqnarray}

\bigskip\noindent
where, as before, velocities were conveniently separated. In the last
constraint, we have not considered the index $\mu=0$ because there is
no other term involving $\buildrel{....}\over A_0$. We have not also
considered the zero components of $T_\mu^{(2)}$ and $T_\mu^{(3)}$
because these components do not constitute independent constraints.

\medskip
With these constraints, we calculate the D.B. involving fields and
their velocities, in the same way we have done in the previous
approximation. The quantization of the present approximation is
expressed by the following commutators:

\begin{eqnarray}
\bigl[\dot A_\mu(\vec x,t),A^\nu(\vec y,t)\bigr]
&=&i\,\alpha\,\delta_\mu^0\delta_0^\nu\,\delta(\vec x-\vec y)\,,
\nonumber\\
\bigl[\ddot A_\mu(\vec x,t),A^\nu(\vec y,t)\bigr]
&=&i\,\alpha\,\delta_\mu^k\delta_0^\nu\,\partial_k\,
\delta(\vec x-\vec y)\,,
\nonumber\\
\bigl[\buildrel{...}\over A_\mu(\vec x,t),A^\nu(\vec y,t)\bigr]
&=&0\,,
\nonumber\\
\bigl[\buildrel{....}\over A_\mu(\vec x,t),A^\nu(\vec y,t)\bigr]
&=&0\,,
\nonumber\\
\bigl[\buildrel{(v)}\over{A_\mu}(\vec x,t),A^\nu(\vec y,t)\bigr]
&=&-\,i\,\delta_\mu^\nu\,\frac{\nabla^6}{m^2}\,\delta(\vec x-\vec y)\,,
\label{3.28}
\end{eqnarray}

\bigskip\noindent
where $\buildrel{(v)}\over{A_\mu}$ stands for five time derivatives
over $A_\mu$. Using the commutators above, one can also show that the
propagator satisfies a similar relation like (3.23) with the operator
that appears in the equation of motion (3.25). This shows that the
commutators above are also consistent relations.

\medskip
Now it is not difficult to infer the commutator relations when all
the terms of the operator $\Box^{-1}$ are taken into account. These
are given by

\begin{eqnarray}
\bigl[\dot A_\mu(\vec x,t),A^\nu(\vec y,t)\bigr]
&=&i\,\alpha\,\delta_\mu^0\delta_0^\nu\,\delta(\vec x-\vec y)\,,
\nonumber\\
\bigl[\ddot A_\mu(\vec x,t),A^\nu(\vec y,t)\bigr]
&=&i\,\alpha\,\delta_\mu^k\delta_0^\nu\partial_k\,
\delta(\vec x-\vec y)\,,
\nonumber\\
\bigl[\buildrel{...}\over {A_\mu}(\vec x,t),A^\nu(\vec y,t)\bigr]
&=&0\,,
\nonumber\\
\bigl[\buildrel{....}\over A_\mu(\vec x,t),A^\nu(\vec y,t)\bigr]
&=&0\,,
\nonumber\\
\bigl[\buildrel{(v)}\over A_\mu(\vec x,t),A^\nu(\vec y,t)\bigr]
&=&0\,,
\nonumber\\
&\vdots&
\nonumber\\
\lim_{n\rightarrow\infty}\Biggl(
\bigl[\buildrel{(2n-1)}\over A_\mu(\vec x,t),A^\nu(\vec y,t)
\bigr]\Biggr)
&=&-\,i\,\delta_\mu^\nu\,\frac{\nabla^{2n}}{m^2}\,
\delta(\vec x-\vec y)\,.
\label{3.29}
\end{eqnarray}

\bigskip
We see in this way that the canonical structure, despite its
nonlocality, is perfectly consistent with the functional procedure,
generating propagators for the vectorial theory which display the
presence of a massive field. To develop the theory furthermore,
trying to construct the Fock space by introducing creation and
annihilation operators for the vectorial fields, this seems to be
nontrivial. This is so because there is no way, if $m$ does not
vanish, to avoid a canonical dependence between the vectorial field
and its derivative of order $2n-1$, in the limit when $n$ goes to
infinity. We may say that the set of equations (\ref{3.29}) shows us
where the nonlocability problem appears in the process of
quantization of these theories.

\vspace{1cm}
\section{Conclusion}

\bigskip
In this work we have considered the quantization of a nonlocal
massive vector gauge invariant field theory, which can be effectively
obtained from a vector-tensor theory with topological coupling. We
have quantized this nonlocal system first by using the BV Lagrangian
functional formalism, where the propagator could be obtained without
major problems.  After that, we have considered its canonical
quantization, where the nonlocalibility becomes a more difficult
problem to be circumvented.  The non-localibility manifests itself
through the canonical independence, at commutator level, between the
gauge field and its derivative of order $n$, in the limit when $n$
goes to infinity. We have shown, however, that a systematic use of
the canonical quantization procedure order by order permitted us to
generate the same Greens functions as those obtained from the
functional formalism. On the other hand, the Fock space structure
seems difficult to be displayed, due to the odd canonical structure
generated by the system. As it would be expected, when the mass
parameter goes to zero, the tensor and vector sector of the theory
decouple, and in the effective vector theory new constraints arise as
a consequence of this limit. We have also shown in a appendix that a
careful analysis of these constraints leads to a canonical structure
that is identical to the usual massless gauge theory.

\medskip
We could argue about the functional Hamiltonian quantization,
due to Batalin, Fradkin and Vilkovisky (BFV) \cite{BFV}, of this
nonlocal system. The use of this formalism here appears to be a
nontrivial task. Even with the procedure of how to circumvent the
infinity number of momenta, we still have an additional problem
because velocities have to be considered as independent canonical
coordinates in higher derivative systems. Consequently, it is
necessary to distinguish in the Hamiltonian path integral formalism
what is a time derivative of a coordinate and what is an independent
coordinate itself \cite{Barc4}. This problem is presently under study
and possible results shall be reported elsewhere \cite{Barc5}.

\vspace{1cm}
\noindent {\bf Acknowledgment:} This work is supported in part by
Conselho Nacional de Desenvolvimento Cient\'{\i}fico e Tecnol\'ogico
- CNPq, Financiadora de Estudos e Projetos - FINEP, and
Funda\c{c}\~ao Universit\'aria Jos\'e Bonif\'acio - FUJB (Brazilian
Research Agencies). 

\vspace{1cm}
\appendix
\renewcommand{\theequation}{A.\arabic{equation}}
\setcounter{equation}{0}
\section*{Appendix A}

\bigskip
In this Appendix, we present some details of the calculation of Eq.
(\ref{3.3}). Considering the Lagrangian (\ref{3.1}), we have for a
general variation of the corresponding action,

\begin{eqnarray}
\delta\int_{t_0}^t d\tau\int d^3\vec x\,\,{\cal L}
&=&-\int_{t_0}^t d\tau\int d^3\vec x\,
\biggl[\,\frac{1}{2}\,\partial_\mu\delta A_\nu\,
\Bigl(1+\frac{m^2}{\Box}\Bigr)\,F^{\mu\nu}
\nonumber\\
&&\phantom{-\int_{t_0}^t d\tau\int d^3\vec x\,\biggl[}
+\frac{1}{2}\,F_{\mu\nu}\,\Bigl(1+\frac{m^2}{\Box}\Bigr)\,
\partial^\mu\delta A^\nu
\nonumber\\
&&\phantom{-\int_{t_0}^t d\tau\int d^3\vec x\,\biggl[}
+\frac{1}{\alpha}\,\bigl(\partial_\mu A^\mu\bigr)\,
\partial_\nu\delta A^\nu\biggr]\,.
\label{A.1}
\end{eqnarray}

\bigskip
\noindent
Let us consider each term of the above expression in a separate way.
The development of the first term leads to

\begin{eqnarray}
&&-\,\frac{1}{2}\,\int_{t_0}^td\tau\int d^3\vec x\,
\partial_\mu\delta A_\nu\,
\Bigl(1+\frac{m^2}{\Box}\Bigr)\,F^{\mu\nu}
\nonumber\\
&&\phantom{-\,\frac{1}{2}\,\int_{t_0}^td\tau}
=-\,\frac{1}{2}\,\int_{t_0}^td\tau\int d^3\vec x\,
\biggl\{\partial_\mu\,\biggl[\delta A_\nu\,
\Bigl(1+\frac{m^2}{\Box}\Bigr)\,F^{\mu\nu}\biggr]
-\delta A_\nu\,\Bigl(1+\frac{m^2}{\Box}\Bigr)\,
\partial_\mu F^{\mu\nu}\biggr\}\,,
\nonumber\\
&&\phantom{-\,\frac{1}{2}\,\int_{t_0}^td\tau}
=-\,\frac{1}{2}\,\int d^3\vec x\,\delta A_\nu\,
\Bigl(1+\frac{m^2}{\Box}\Bigr)\,F^{0\nu}
\nonumber\\
&&\phantom{-\,\frac{1}{2}\,\int_{t_0}^td\tau=}
+\frac{1}{2}\,\int_{t_0}^td\tau\int d^3\vec x\,
\delta A_\nu\,\Bigl(1+\frac{m^2}{\Box}\Bigr)\,
\partial_\mu F^{\mu\nu}\,.
\label{A.2}
\end{eqnarray}

\bigskip
\noindent
For the second term, we have

\begin{eqnarray}
&&-\,\frac{1}{2}\,\int_{t_0}^td\tau\int d^3\vec x\,F_{\mu\nu}\,
\Bigl(1+\frac{m^2}{\Box}\Bigr)\,\partial^\mu\delta A^\nu
\nonumber\\
&&\phantom{-\,\frac{1}{2}\,\int_{t_0}^td\tau}
=-\frac{1}{2}\,\int_{t_0}^td\tau\int d^3\vec x\,
\biggl\{\partial^\mu\,\biggl[F_{\mu\nu}
\Bigl(1+\frac{m^2}{\Box}\Bigr)\,\delta A^\nu\,\biggr]
-\partial^\mu F_{\mu\nu}\,\Bigl(1+\frac{m^2}{\Box}\Bigr)\,
\delta A^\nu\biggr\}\,,
\nonumber\\
&&\phantom{-\,\frac{1}{2}\,\int_{t_0}^td\tau}
=-\,\frac{1}{2}\,\int d^3\vec x\,F_{0\nu}
\Bigl(1+\frac{m^2}{\Box}\Bigr)\,\delta A^\nu
+\frac{1}{2}\,\int_{t_0}^td\tau\int d^3\vec x\,
\partial^\mu F_{\mu\nu}\,\Bigl(1+\frac{m^2}{\Box}\Bigr)\,
\delta A_\nu\,,
\nonumber\\
&&\phantom{-\,\frac{1}{2}\,\int_{t_0}^td\tau}
=-\,\frac{1}{2}\int d^3\vec x\,F_{0\nu}\,\delta A^\nu
+\frac{1}{2}\,\int_{t_0}^td\tau\int d^3\vec x\,
\partial^\mu F_{\mu\nu}\,\delta A^\nu
\nonumber\\
&&\phantom{-\,\frac{1}{2}\,\int_{t_0}^td\tau=}
+\,\frac{m^2}{2}\int d^3\vec x\,F_{0\nu}\,
\Bigl(\frac{1}{\nabla^2}+\frac{\partial_t^2}{\nabla^4}
+\cdots\Bigr)\,\delta A^\nu
\nonumber\\
&&\phantom{-\,\frac{1}{2}\,\int_{t_0}^td\tau=}
-\,\frac{m^2}{2}\,\int_{t_0}^td\tau\int d^3\vec x\,
\partial^\mu F_{\mu\nu}\,
\Bigl(\frac{1}{\nabla^2}+\frac{\partial_t^2}{\nabla^4}
+\cdots\Bigr)\,\delta A^\nu\,,
\label{A.3}
\end{eqnarray}

\bigskip\noindent
where we have used the expansion (\ref{3.2}). We observe that in the
last term of the expression above, there is an integration over time
and an infinite number of time derivatives acting over $\delta
A^\nu$.  It is necessary some care to deal with these terms. Let us
consider some of them isolately

\begin{eqnarray}
&&-\,\frac{m^2}{2}\,\int_{t_0}^td\tau\int d^3\vec x\,
\partial^\mu F_{\mu\nu}\,\frac{\partial_t^2}{\nabla^4}\,
\delta A^\nu
\nonumber\\
&&\phantom{-\,\frac{m^2}{2}\,\int_{t_0}^td\tau}
=-\,\frac{m^2}{2}\,\int_{t_0}^td\tau\int d^3\vec x\,
\biggl\{\partial_t\Bigl[\frac{\partial^\mu}{\nabla^4}\,
F_{\mu\nu}\,\partial_t\,\delta A^\nu\Bigr]
-\frac{\partial_t\partial^\mu}{\nabla^4}\,F_{\mu\nu}\,
\partial_t\delta A^\nu\biggr\}\,,
\nonumber\\
&&\phantom{-\,\frac{m^2}{2}\,\int_{t_0}^td\tau}
=-\,\frac{m^2}{2}\int d^3\vec x\,
\frac{\partial^\mu}{\nabla^4}\,F_{\mu\nu}\,\delta\dot A^\nu
\nonumber\\
&&\phantom{-\,\frac{m^2}{2}\,\int_{t_0}^td\tau=}
+\,\frac{m^2}{2}\,\int_{t_0}^td\tau\int d^3\vec x\,
\biggl\{\partial_t\Bigl[\frac{\partial_t\partial^\mu}{\nabla^4}\,
F_{\mu\nu}\,\delta A^\nu\Bigr]
-\frac{\partial_t^2\partial^\mu}{\nabla^4}\,F_{\mu\nu}\,
\delta A^\nu\biggr\}\,,
\nonumber\\
&&\phantom{-\,\frac{m^2}{2}\,\int_{t_0}^td\tau}
=-\,\frac{m^2}{2}\int d^3\vec x\,
\frac{\partial^\mu}{\nabla^4}\,F_{\mu\nu}\,\delta\dot A^\nu
+\,\frac{m^2}{2}\int d^3\vec x\,
\frac{\partial^\mu}{\nabla^4}\,\dot F_{\mu\nu}\,\delta A^\nu
\nonumber\\
&&\phantom{-\,\frac{m^2}{2}\,\int_{t_0}^td\tau=}
-\,\frac{m^2}{2}\int_{t_0}^td\tau\int d^3\vec x\,
\frac{\partial^\mu}{\nabla^4}\,\ddot F_{\mu\nu}\,\delta A^\nu\,.
\label{A.4}
\end{eqnarray}

\bigskip\noindent
In a similar way, we would have for the next term

\begin{eqnarray}
&&-\,\frac{m^2}{2}\,\int_{t_0}^td\tau\int d^3\vec x\,
\partial^\mu F_{\mu\nu}\,\frac{\partial_t^4}{\nabla^6}\,\delta A^\nu
\phantom{-\,\frac{m^2}{2}\,\int_{t_0}^td\tau}
\nonumber\\
&&\phantom{-\,\frac{m^2}{2}\,\int_{t_0}^td\tau}
=-\,\frac{m^2}{2}\int d^3\vec x\,
\frac{\partial^\mu}{\nabla^6}\,F_{\mu\nu}\,
\delta\buildrel{...}\over A^\nu
+\,\frac{m^2}{2}\int d^3\vec x\,
\frac{\partial^\mu}{\nabla^6}\,\dot F_{\mu\nu}\,
\delta\ddot A^\nu
\nonumber\\
&&\phantom{-\,\frac{m^2}{2}\,\int_{t_0}^td\tau=}
-\,\frac{m^2}{2}\int d^3\vec x\,
\frac{\partial^\mu}{\nabla^6}\,\ddot F_{\mu\nu}\,
\delta\dot A^\nu
+\,\frac{m^2}{2}\int d^3\vec x\,
\frac{\partial^\mu}{\nabla^6}\,\buildrel{...}\over F_{\mu\nu}\,
\delta A^\nu
\nonumber\\
&&\phantom{-\,\frac{m^2}{2}\,\int_{t_0}^td\tau=}
-\,\frac{m^2}{2}\int_{t_0}^td\tau\int d^3\vec x\,
\frac{\partial_t^4}{\nabla^6}\,\partial^\mu F_{\mu\nu}\,
\delta A^\nu\,,
\label{A.5}
\end{eqnarray}

\bigskip\noindent
and so on. Introducing these results into the initial expression
(\ref{A.3}), we obtain

\begin{eqnarray}
&&-\,\frac{1}{2}\,\int_{t_0}^td\tau\int d^3\vec x\,F_{\mu\nu}\,
\Bigl(1+\frac{m^2}{\Box}\Bigr)\,\partial^\mu\delta A^\nu
\nonumber\\
&&\phantom{-\,\frac{1}{2}\,\int_{t_0}^td\tau}
=\frac{1}{2}\,\int_{t_0}^td\tau\int d^3\vec x\,
\Bigl(1+\frac{m^2}{\Box}\Bigr)\,\partial^\mu\,
F_{\mu\nu}\,\delta A^\nu
\nonumber\\
&&\phantom{-\,\frac{1}{2}\,\int_{t_0}^td\tau}
+\,\frac{1}{2}\,\int d^3\vec x\,
\Bigl(-F_{0\nu}+m^2\,\frac{1}{\nabla^2}\,F_{0\nu}
+m^2\,\frac{\partial^\mu}{\nabla^4}\,\dot F_{\mu\nu}
+\dots\Bigr)\delta A^\nu
\nonumber\\
&&\phantom{-\,\frac{1}{2}\,\int_{t_0}^td\tau}
-\,\frac{m^2}{2}\,\int d^3\vec x\,
\Bigl(\frac{\partial^\mu}{\nabla^4}\,F_{\mu\nu}
+\frac{\partial^\mu}{\nabla^6}\,\ddot F_{\mu\nu}
+\frac{\partial^\mu}{\nabla^8}\,\buildrel{....}\over F_{\mu\nu}
+\cdots\Bigr)\delta\dot A^\nu
\nonumber\\
&&\phantom{-\,\frac{1}{2}\,\int_{t_0}^td\tau}
+\,\frac{m^2}{2}\,\int d^3\vec x\,
\Bigl(\frac{1}{\nabla^4}\,F_{0\nu}
+\frac{\partial^\mu}{\nabla^6}\,\dot F_{\mu\nu}
+\frac{\partial^\mu}{\nabla^8}\,\buildrel{...}\over F_{\mu\nu}
+\cdots\Bigr)\delta\ddot A^\nu
\nonumber\\
&&\phantom{-\,\frac{1}{2}\,\int_{t_0}^td\tau}
-\frac{m^2}{2}\,\int d^3\vec x\,
\Bigl(\frac{\partial^\mu}{\nabla^6}\,F_{\mu\nu}
+\frac{\partial^\mu}{\nabla^8}\,\ddot F_{\mu\nu}
+\frac{\partial^\mu}{\nabla^{10}}\,\buildrel{....}\over F_{\mu\nu}
+\cdots\Bigr)\delta\buildrel{...}\over A^\nu
\nonumber\\
&&\phantom{-\,\frac{1}{2}\,\int_{t_0}^td\tau}
+\cdots
\label{A.6}
\end{eqnarray}

\bigskip\noindent
We notice in the expression above that some terms can be put together
to reobtain the nonlocal operator $\Box^{-1}$. For example,

\begin{eqnarray}
&&\frac{1}{\nabla^2}\,F_{0\nu}
+\frac{\partial^\mu}{\nabla^4}\,\dot F_{\mu\nu}
+\frac{\partial^\mu}{\nabla^6}\,\buildrel{...}\over F_{\mu\nu}
+\cdots
\nonumber\\
&&\phantom{\frac{1}{\nabla^2}\,F_{0\nu}}
=\Bigl(\frac{1}{\nabla^2}
+\frac{\partial^2_t}{\nabla^4}+\cdots\Bigr)\,F_{0\nu}
+\Bigl(\frac{1}{\nabla^4}
+\frac{\partial^2_t}{\nabla^6}+\cdots\Bigr)\,
\partial^i\dot F_{i\nu}
\nonumber\\
&&\phantom{\frac{1}{\nabla^2}\,F_{0\nu}}
=-\,\frac{1}{\Box}\,F_{0\nu}
-\frac{1}{\nabla^2}\frac{\partial^i}{\Box}\dot F_{i\nu}\,,
\label{A.7}\\
\nonumber\\
\nonumber\\
&&\frac{\partial^\mu}{\nabla^4}\,F_{\mu\nu}
+\frac{\partial^\mu}{\nabla^6}\,\ddot F_{\mu\nu}
+\frac{\partial^\mu}{\nabla^8}\buildrel{....}\over F_{\mu\nu}
+\cdots
\nonumber\\
&&\phantom{\frac{1}{\nabla^2}\,F_{0\nu}}
=-\,\frac{1}{\nabla^2}\frac{\partial^\mu}{\Box}\,F_{\mu\nu}\,.
\label{A.8}
\end{eqnarray}

\bigskip\noindent
and so on. Using these results into (\ref{A.6}), we obtain the final
form of the second term of (\ref{A.1}).

\begin{eqnarray}
&&-\,\frac{1}{2}\,\int_{t_0}^td\tau\int d^3\vec x\,F_{\mu\nu}\,
\Bigl(1+\frac{m^2}{\Box}\Bigr)\,\partial^\mu\delta A^\nu
\nonumber\\
&&\phantom{-\,\frac{1}{2}\,\int_{t_0}^td\tau}
=\frac{1}{2}\,\int_{t_0}^td\tau\int d^3\vec x\,
\Bigl(1+\frac{m^2}{\Box}\Bigr)\,\partial^\mu\,
F_{\mu\nu}\,\delta A^\nu
\nonumber\\
&&\phantom{-\,\frac{1}{2}\,\int_{t_0}^td\tau}
-\,\frac{1}{2}\,\int d^3\vec x\,
\biggl[\Bigl(1+\frac{m^2}{\Box}\Bigr)\,F_{0\nu}
+\frac{m^2}{\nabla^2}\frac{\partial^i}{\Box}\,F_{i\nu}
\biggr]\,\delta A^\nu
\nonumber\\
&&\phantom{-\,\frac{1}{2}\,\int_{t_0}^td\tau}
+\,\frac{m^2}{2}\,\int d^3\vec x\,
\frac{1}{\nabla^2}\frac{\partial^\mu}{\Box}\,
F_{\mu\nu}\,\delta\dot A^\nu
\nonumber\\
&&\phantom{-\,\frac{1}{2}\,\int_{t_0}^td\tau}
-\,\frac{m^2}{2}\,\int d^3\vec x\,
\Bigl(\frac{1}{\nabla^2}\frac{1}{\Box}\,F_{0\nu}
+\frac{1}{\nabla^4}\frac{\partial^i}{\Box}\,\dot F_{i\nu}\Bigr)\,
\delta\ddot A^\nu
\nonumber\\
&&\phantom{-\,\frac{1}{2}\,\int_{t_0}^td\tau}
+\,\frac{m^2}{2}\,\int d^3\vec x\,
\frac{1}{\nabla^4}\frac{\partial^\mu}{\Box}\,
F_{\mu\nu}\,\delta\buildrel{...}\over A^\nu
\nonumber\\
&&\phantom{-\,\frac{1}{2}\,\int_{t_0}^td\tau}
+\cdots
\label{A.9}
\end{eqnarray}

\bigskip\noindent
We finally consider the last term of expression (\ref{A.1}).

\begin{eqnarray}
&&-\,\frac{1}{\alpha}\,\int_{t_0}^td\tau\int d^3\vec x\,
\partial_\mu A^\mu\,\partial_\nu\delta A^\nu
\nonumber\\
&&\phantom{-\,\frac{1}{2}\,\int_{t_0}^td\tau}
=-\,\frac{1}{\alpha}\,\int_{t_0}^td\tau\int d^3\vec x\,
\Bigl[\partial_\nu\bigl(\partial_\mu A^\mu\delta A^\nu\bigr)
-\partial_\nu\partial_\mu A^\mu\delta A^\nu\Bigr]\,,
\nonumber\\
&&\phantom{-\,\frac{1}{2}\,\int_{t_0}^td\tau}
=-\,\frac{1}{\alpha}\,\int d^3\vec x\,
\partial_\mu A^\mu\delta A^0
+\frac{1}{\alpha}\,\int_{t_0}^td\tau\int d^3\vec x\,
\partial_\nu\partial_\mu A^\mu\delta A^\nu\,.
\label{A.10}
\end{eqnarray}

\bigskip\noindent
Introducing the results given by expressions (\ref{A.2}), (\ref{A.9})
and (\ref{A.10}) into the initial expression (\ref{A.1}), the
equation (\ref{3.3}) is obtained.

\vspace{1cm}
\appendix
\renewcommand{\theequation}{B.\arabic{equation}}
\setcounter{equation}{0}
\section*{Appendix B}

\bigskip
In this appendix, we calculate the inverse of the matrix
(\ref{3.17}).  First we notice it has the following block structure

\begin{equation}
S=\left(\begin{array}{ccc}
{\left(\begin{array}{c}
1\times1\\
\end{array}\right)}
&{\left(\begin{array}{ccccc}&&1\times4&&\\
\end{array}\right)}
&{\left(\begin{array}{ccc}&1\times3&\\
\end{array}\right)}\\
&&\\
{\left(\begin{array}{c}
\\
\\
4\times1\\
\\
\\
\end{array}\right)}
&{\left(\begin{array}{ccccc}
&&&&\\
&&&&\\
&&4\times4&&\\
&&&&\\
&&&&\\
\end{array}\right)}
&{\left(\begin{array}{ccc}
&&\\
&&\\
&4\times3&\\
&&\\
&&\\
\end{array}\right)}\\
&&\\
{\left(\begin{array}{c}
\\
3\times1\\
\\
\end{array}\right)}
&{\left(\begin{array}{ccccc}
&&&&\\
&&3\times4&&\\
&&&&\\
\end{array}\right)}
&{\left(\begin{array}{ccc}
&&\\
&3\times3&\\
&&\\
\end{array}\right)}\\
\end{array}\right)\,.
\label{B.1}
\end{equation}

\bigskip\noindent
Of course, since the inverse $S^{-1}$ has the same block structure,
we consider it is generically given by

\begin{equation}
S^{-1}=\left(\begin{array}{ccc}
A&B^\rho&C^k\\
D_\nu&E_\nu^\rho&F_\nu^k\\
G_j&H_j^\rho&I_j^k\\
\end{array}\right)\,.
\label{B.2}
\end{equation}

\bigskip\noindent
We then must have

\begin{equation}
\int d^3\vec y\,\, S(\vec x,\vec y)S^{-1}(\vec y-\vec z)
=\left(\begin{array}{ccc}
1&0&0\\
0&\delta_\mu^\rho&0\\
0&0&\delta_j^k\\
\end{array}\right)\,\delta(\vec x-\vec z)\,.
\label{B.3}
\end{equation}

\bigskip\noindent
The combination of Eqs. (\ref{3.17}), (\ref{B.2}) and (\ref{B.3}) gives us
the following set of equations (after integrating over the
intermediary variable $\vec y$ and summing on mudding indices):

\begin{eqnarray}
&&\Bigl(\frac{m^2}{\nabla^2}+\frac{1}{\alpha}\Bigr)\,D_0
+\frac{m^2}{\nabla^4}\,\partial^jG_j=\delta(\vec x-\vec z)\,,
\nonumber\\
&&\Bigl(\frac{m^2}{\nabla^2}+\frac{1}{\alpha}\Bigr)\,E_0^\rho
+\frac{m^2}{\nabla^4}\,\partial^jH_j^\rho=0\,,
\nonumber\\
&&\Bigl(\frac{m^2}{\nabla^2}+\frac{1}{\alpha}\Bigr)\,F_0^k
+\frac{m^2}{\nabla^4}\,\partial^jI_j^k=0
\label{B.4}\\
\nonumber\\
\nonumber\\
&&\delta_\mu^0\,\Bigl(\frac{m^2}{\nabla^2}+\frac{1}{\alpha}\Bigr)\,A
+\delta_\mu^0\,\frac{m^2}{\nabla^4}\partial_jD^j
+\delta_\mu^j\,\frac{m^2}{\nabla^4}\partial_jD^0
-\delta_\mu^j\,\frac{m^2}{\nabla^4}G_j=0\,,
\nonumber\\
&&\delta_\mu^0\,\Bigl(\frac{m^2}{\nabla^2}
+\frac{1}{\alpha}\Bigr)\,B^\rho
+\delta_\mu^0\,\frac{m^2}{\nabla^4}\partial^jE_j^\rho
+\delta_\mu^j\,\frac{m^2}{\nabla^4}\partial_jE_0^\rho
-\delta_\mu^j\,\frac{m^2}{\nabla^4}H_j^\rho
=\delta_\mu^\rho\,\delta(\vec x-\vec y)\,,
\nonumber\\
&&\delta_\mu^0\,\Bigl(\frac{m^2}{\nabla^2}
+\frac{1}{\alpha}\Bigr)\,C^k
+\delta_\mu^0\,\frac{m^2}{\nabla^4}\partial^jF_j^k
+\delta_\mu^j\,\frac{m^2}{\nabla^4}\partial_jF_0^k
-\delta_\mu^j\,\frac{m^2}{\nabla^4}I_j^k=0
\label{B.5}\\
\nonumber\\
\nonumber\\
&&\partial_iA-D_i=0\,,
\nonumber\\
&&\partial_iB^\rho-E_i^\rho=0
\nonumber\\
&&\partial_iC^k-F_i^k=\frac{\nabla^4}{m^2}\,\delta_i^k\,
\delta(\vec x-\vec z)\,,
\label{B.6}\\
\nonumber\\
\nonumber\\
&&\Bigl(\frac{m^2}{\nabla^2}+\frac{1}{\alpha}\Bigr)\,A
+\frac{m^2}{\nabla^4}\,\partial^jD_j=0\,,
\nonumber\\
&&\Bigl(\frac{m^2}{\nabla^2}+\frac{1}{\alpha}\Bigr)\,B^0
+\frac{m^2}{\nabla^4}\,\partial^jE_j^0
=-\,\delta(\vec x-\vec z)\,,
\nonumber\\
&&\Bigl(\frac{m^2}{\nabla^2}+\frac{1}{\alpha}\Bigr)\,B^k
+\frac{m^2}{\nabla^4}\,\partial^jE_j^k=0
\nonumber\\
&&\Bigl(\frac{m^2}{\nabla^2}+\frac{1}{\alpha}\Bigr)\,C^k
+\frac{m^2}{\nabla^4}\,\partial^jF_j^k=0
\nonumber\\
&&\partial_iE_0^0-H_i^0=0\,,
\nonumber\\
&&\partial_iE_0^k-H_i^k
=-\,\frac{\nabla^4}{m^2}\delta_i^k\,\delta(\vec x-\vec z)\,,
\nonumber\\
&&\partial_iD_0-G_i=0\,,
\nonumber\\
&&\partial_iF_0^k-I_i^k=0\,.
\label{B.20}
\end{eqnarray}

\bigskip\noindent
The inverse $S^{-1}$ is obtained by solving these equations. This is
just a matter of algebraic work and the solution is

\begin{eqnarray}
A&=&0\,,
\nonumber\\
B^0&=&-\,\alpha\,\delta(\vec x-\vec z)\,,
\nonumber\\
B^k&=&0\,,
\nonumber\\
C^k&=&\alpha\,\partial^k\delta(\vec x-\vec z)\,,
\nonumber\\
D_0&=&\alpha\,\delta(\vec x-\vec z)\,,
\nonumber\\
D_i&=&0\,,
\nonumber\\
E_0^0&=&0\,,
\nonumber\\
E_i^0&=&-\,\alpha\,\partial_i\,\delta(\vec x-\vec z)\,,
\nonumber\\
E_0^k&=&-\,\alpha\,\partial^k\delta(\vec x-\vec z)\,,
\nonumber\\
E_i^k&=&0\,,
\nonumber\\
F_0^k&=&0\,,
\nonumber\\
F_i^k&=&\Bigl(\alpha\,\partial_i\partial^k
-\delta_i^k\frac{\nabla^4}{m^2}\Bigr)\,\delta(\vec x-\vec z)\,,
\nonumber\\
G_j&=&\alpha\,\partial_j\delta(\vec x-\vec z)\,,
\nonumber\\
H_j^0&=&0\,,
\nonumber\\
H_j^k&=&-\,\Bigl(\alpha\,\partial_j\partial^k
-\delta_j^k\frac{\nabla^4}{m^2}\Bigr)\,\delta(\vec x-\vec z)\,,
\nonumber\\
I_j^k&=&0\,.
\label{B.21}
\end{eqnarray}

\bigskip\noindent
These are the elements of the matrix $S^{-1}$ given by (\ref{3.18}).

\vspace{1cm}
\appendix
\renewcommand{\theequation}{C.\arabic{equation}}
\setcounter{equation}{0}
\section*{Appendix C}

\bigskip
When we take the limit $m\rightarrow0$ in expressions (\ref{3.5}), we
get the following expressions for the momenta

\begin{eqnarray}
\pi_\nu&=&-\,F_{0\nu}
-\frac{1}{\alpha}\,\eta_{0\nu}\,\partial_\mu A^\mu\,,
\nonumber\\
\pi_\nu^{(1)}&=&0\,.
\label{C.1}
\end{eqnarray}

\bigskip\noindent
The remaining momenta (which are all zero) do not make sense to be
considered because in the limit $m\rightarrow0$ the system does not
have infinite derivatives anymore. We observe that relations
(\ref{C.1}) are constrains. So, the commutators cannot come from the
P.B. of $A_\mu$ and $\dot A^\nu$, that is actually zero, but from the
Dirac one. Let us calculate the D.B. of $A_\mu$ and $\dot A_\nu$.
First, we need the P.B. matrix of the constraints. We denote these
constraints by

\begin{eqnarray}
T_{1\mu}&=&\pi_\mu+\,F_{0\mu}
+\frac{1}{\alpha}\,\eta_{0\mu}\,\partial_\nu A^\nu\,,
\nonumber\\
T_{2\mu}&=&\pi_\mu^{(1)}\,.
\label{C.2}
\end{eqnarray}

\bigskip\noindent
Thus
\begin{eqnarray}
\bigl(S^{\mu\nu}\bigr)
&=&\left(\begin{array}{cc}
\{T_1^\mu,T_1^\nu\}&\{T_1^\mu,T_2^\nu\}\\
&\\
\{T_2^\mu,T_1^\nu\}&\{T_2^\mu,T_2^\nu\}\\
\end{array}\right)
\nonumber\\
\nonumber\\
&=&\left(\begin{array}{cc}
0&1\\
-1&0\\
\end{array}\right)\,
\Bigl(\eta^{\mu\nu}+\frac{1-\alpha}{\alpha}\,
\eta^{0\mu}\eta^{0\nu}\Bigr)\,
\delta(\vec x-\vec y)\,.
\label{C.3}
\end{eqnarray}

\bigskip\noindent
To calculate the D.B. we need the inverse of the matrix above. This
is give by

\begin{equation}
\bigl(S_{\mu\nu}\bigr)^{-1}=\left(\begin{array}{cc}
0&-1\\
1&0\\
\end{array}\right)\,
\Bigl(\eta_{\mu\nu}+(\alpha-1)\,\eta_{0\mu}\eta_{0\nu}\Bigr)\,
\delta(\vec x-\vec y)\,.
\label{C.4}
\end{equation}

\bigskip\noindent
Now, the D.B. can be directly calculated. The result is

\begin{equation}
\bigl\{A_\mu(\vec x,t),\dot A^\nu(\vec y,t)\bigr\}
=-\,\Bigl(\delta_\mu^\nu
+(\alpha-1)\,\eta_{0\mu}\delta_0^\nu\Bigr)\,
\delta(\vec x-\vec y)\,.
\label{C.5}
\end{equation}

\bigskip\noindent
The commutator between $A_i$ and $\dot A^j$ can be directly obtained
from the D.B. above and it is actually non zero when $m\rightarrow0$,
what makes consistent the procedure we are developing.

\vspace{1cm}
\appendix
\renewcommand{\theequation}{D.\arabic{equation}}
\setcounter{equation}{0}
\section*{Appendix D}

\bigskip
Considering that

\begin{equation}
T\Bigl(A^\nu(x)A^\rho(x^\prime)\Bigr)
=\theta(t-t^\prime)\,A^\nu(x)A^\rho(x^\prime)
+\theta(t^\prime-t)\,A^\rho(x^\prime)A^\nu(x)
\label{D.1}
\end{equation}

\bigskip\noindent
and using the commutators given by expression (\ref{3.21}) we can
obtain the following relations

\begin{eqnarray}
&&\frac{\partial^2}{\partial t^2}\,
T\Bigl(A^\nu(x)A^\rho(x^\prime)\Bigr)
=i\,\alpha\,\delta_0^\nu\eta^{0\rho}\,\delta(x-x^\prime)
+T\Bigl(\ddot A^\nu(x)A^\rho(x^\prime)\Bigr)\,,
\nonumber\\
&&\nabla^2\,T\Bigl(A^\nu(x)A^\rho(x^\prime)\Bigr)
=T\Bigl(\nabla^2A^\nu(x)A^\rho(x^\prime)\Bigr)\,,
\nonumber\\
&&\Box\,T\Bigl(A^\nu(x)A^\rho(x^\prime)\Bigr)
=i\,\alpha\,\delta_0^\nu\eta^{0\rho}\,\delta(x-x^\prime)
+T\Bigl(\Box A^\nu(x)A^\rho(x^\prime)\Bigr)\,,
\nonumber\\
&&\frac{1}{\nabla^2}\,\Box\,T\Bigl(A^\nu(x)A^\rho(x^\prime)\Bigr)
=i\,\alpha\,\delta_0^\nu\eta^{0\rho}\,
\frac{1}{\nabla^2}\,\delta(x-x^\prime)
+T\Bigl(\frac{1}{\nabla^2}\Box A^\nu(x)A^\rho(x^\prime)\Bigr)\,,
\nonumber\\
&&\frac{\partial_t^2}{\nabla^2}\,\Box\,
T\Bigl(A^\nu(x)A^\rho(x^\prime)\Bigr)
=i\,\Bigl[\alpha\,\delta_0^\nu\eta^{0\rho}
\frac{1}{\nabla^2}\,\Box+\alpha\,\eta^{\nu i}\delta_0^\rho\,
\frac{\partial_t\partial_i}{\nabla^2}
\nonumber\\
&&\phantom{\frac{\partial_t^2}{\nabla^2}\Box\,T(A^\nu(x)}
-\,\eta^{\nu\rho}\frac{\nabla^2}{m^2}\Bigr]\,\delta(x-x^\prime)
+T\Bigl(\frac{\partial_t^2}{\nabla^2}
\Box A^\nu(x)A^\rho(x^\prime)\Bigr)\,,
\nonumber\\
&&\partial_\mu\partial_\nu T\Bigl(A^\nu(x)A^\rho(x^\prime)\Bigr)
=i\,\alpha\,\delta_\mu^0\delta_0^\rho\,\delta(x-x^\prime)
+T\Bigl(\partial_\mu\partial_\nu A^\nu(x)A^\rho(x^\prime)\Bigr)\,,
\nonumber\\
&&\frac{\partial_\mu\partial_\nu\partial^2_t}{\nabla^4}\,
T(A^\nu(x)A^\rho(x^\prime))
=i\,\Bigl[\alpha\,\delta_\mu^0\delta_0^\rho\,
\frac{\partial_t^2}{\nabla^4}
+\alpha\,\delta_\mu^i\delta_0^\rho\,
\frac{\partial_i\partial_t}{\nabla^4}
-\delta_\mu^0\delta^\rho_0\,\frac{1}{m^2}
\nonumber\\
&&\phantom{\frac{\partial_t^2}{\nabla^2}\Box\,T(A^\nu(x)}
-\alpha\,\delta_\mu^0\delta^\rho_0\,\frac{1}{\nabla^2}
\Bigr]\,\delta(x-x^\prime)
+T\Bigl(\frac{\partial_\mu\partial_\nu\partial_t^2}{\nabla^4}\,
A^\nu(x)A^\rho(x^\prime)\Bigr)\,.
\label{D.2}
\end{eqnarray}

\bigskip\noindent
Using the relations above in the left side of Eq. (\ref{3.23}) we can
show that the identity is actually satisfied.

\end{document}